\documentclass[superscriptaddress,prl,twocolumn,amsmath,amssymb,floatfix,showpacs]{revtex4-1}
\usepackage{graphicx}
\usepackage{bm}
\usepackage{amssymb}
\usepackage{color}
\usepackage{epsfig}
\usepackage{subfigure}
\usepackage{graphicx}
\usepackage{tikz}

\usepackage{amsmath}
\usepackage{hyperref}
\usepackage{bbold}
\hypersetup{
    pdfstartview={FitH},
   }
\newcommand{\be}{\begin{equation}}
\newcommand{\ee}{\end{equation}}
\begin{document}


\title{Unsupervised learning of correlated quantum dynamics on disordered lattices}

\author{Miri Kenig}
\affiliation{Department of Condensed Matter Physics, School of Physics and Astronomy, Tel-Aviv University, Tel Aviv 6997801, Israel}
\author{Yoav Lahini}
\affiliation {Department of Condensed Matter Physics, School of Physics and Astronomy, Tel-Aviv University, Tel Aviv 6997801, Israel}

\begin{abstract}
Quantum particles co-propagating on disordered lattices develop complex non-classical correlations due to an interplay between quantum statistics, inter-particle interactions, and disorder. Here we present a deep learning algorithm based on Generative Adversarial Networks, capable of learning these correlations and identifying the physical control parameters in a completely unsupervised manner. After one-time training on a data set of unlabeled examples, the algorithm can generate, without further calculations, a much larger number of unseen yet physically correct new examples. Furthermore, the knowledge distilled in the algorithm's latent space identifies disorder as the relevant control parameter. This allows post-training tuning of the level of disorder in the generated samples to values the algorithm was not explicitly trained on. Finally, we show that a trained network can accelerate the learning of new, more complex problems. These results demonstrate the ability of neural networks to learn the rules of correlated quantum dynamics in an unsupervised manner and offer a route to their use in quantum simulations and computation.  

\end{abstract}

\maketitle
Quantum walks, the quantum analogs of random walks, are fundamental processes describing the evolution of an initially localized quantum particle on a lattice potential \cite{aharonov1993quantum, farhi1998quantum, kempe2003quantum, mulken2011continuous, venegas2012quantum}. As a quantum particle can be in superposition, its evolution is governed by the interference of all the possible paths it can take across the lattice. As a result, quantum walks exhibit several differences compared to classical walks. On periodic potentials they propagate faster: ballistically rather than diffusively \cite{varbanov2008hitting}. On disordered lattices, the quantum walker may come to a complete halt – a phenomenon known as Anderson Localization \cite{PhysRev.109.1492, schwartz2007transport, lahini2008anderson, harris2017quantum}. While single-particle quantum walks can be described as wave interference phenomena, \cite{do2005experimental,perets2008realization,broome2010discrete, harris2017quantum}, non-classical effects emerge when several indistinguishable quantum particles propagate on the lattice simultaneously \cite{bromberg2009quantum, lahini_Qanderson,peruzzo2010quantum, lahini_2particles_PhysRevA.86.011603, childs2013universal, weitenberg2011single, preiss2015strongly}. Here, the collective evolution is governed by the interference of all possible \emph{multi-particle} processes. As a result of quantum statistics, the propagating particles develop non-classical correlations, or dependencies, between their positions as they propagate across the lattice \cite{bromberg2009quantum, peruzzo2010quantum, aaronson2011computational} (see Figure 1). While this is true even when the particles do not interact with each other, interactions introduce additional complexity, reflected in the growth of the Hilbert space required to account for all possible outcomes \cite{lahini_2particles_PhysRevA.86.011603,preiss2015strongly}. 

The ability to shape these high dimensional correlations by controlling the lattice potential, the interactions, or other parameters generated broad interest in applying multi-particle quantum walks for various computation and information processing tasks. These include quantum computation \cite{childs2013universal, lahini2018quantum}, preparing and manipulating quantum states \cite{gao2017efficient,huang2017neural}, devising search algorithm \cite{shenvi2003quantum, childs2004spatial}, implementing perfect state transfer \cite{kay2006perfect}, exploring topological phases \cite{kitagawa2010exploring,panahiyan2019simulation, kraus2012topological, PhysRevLett.121.100502}, evaluating Boolean formulas \cite{farhi1998quantum, ambainis2010any} and more. This interest is further motivated by the recent experimental realizations of pristine quantum-walk systems using photons \cite{do2005experimental,perets2008realization,broome2010discrete,bromberg2009quantum, rohde2011multi,peruzzo2010quantum,aspuru2012photonic}, trapped ions \cite{zahringer2010realization,schmitz2009quantum}, superconducting qubits \cite{flurin2017observing,yan2019strongly, gong2021quantum} , and ultra-cold atoms \cite{karski2009quantum, weitenberg2011single, fukuhara2013microscopic,preiss2015strongly}, which allow exquisite control of the number and initial positions of the particles, of their interactions and of the lattice potential. Yet, the computational cost of multi-particle quantum walk increases sharply with the number of walkers \cite{aaronson2011computational}, placing a limit on the speed and efficiency of simulations and inverse design calculations \cite{lahini2018quantum}. This is particularly true for problems considering correlated dynamics on disordered lattices, which usually require statistical averages over many realizations of disorder.   

\begin{figure*}
\includegraphics[width=1\textwidth]{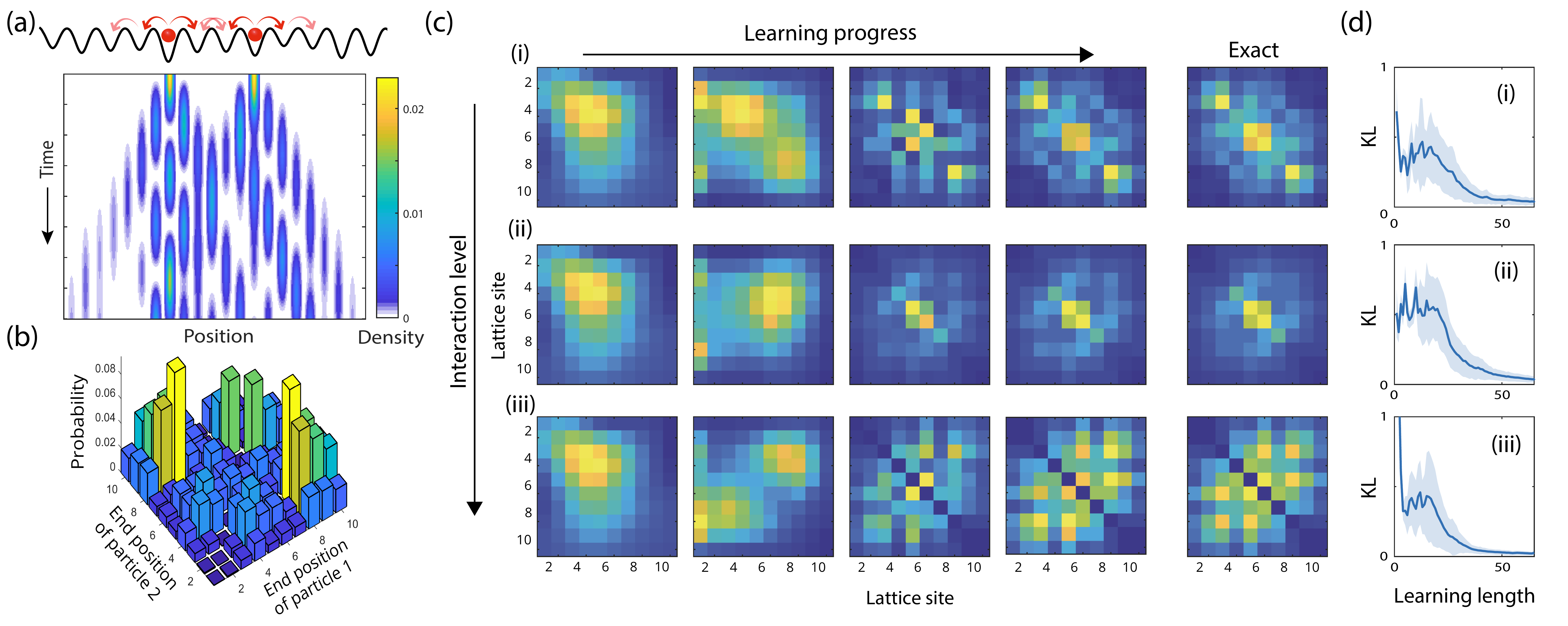}
\caption{(a) an illustration of correlated, two particle quantum walk on a disordered lattice, showing the evolution of the particle density. (b) An example of the resulting two-particle correlation function – the positions of the two particles at the end of the propagation depend on each other. (c) Improvement in generating physically correct correlations as learning progresses, compared to exact calculation (last column) for: (i) $\Gamma=0$, no interactions; (ii) $\Gamma=3$, intermediate interactions; (iii) $\Gamma=1000$, strong interactions. Results are for $\left|\psi_{initial}\right\rangle=a_{5}^{\dagger}a_{6}^{\dagger}\left|0\right\rangle $, averaged over 1000 realizations of disorder. (d) The physical validity of the generated correlation functions as a function of the learning length, quantified using the Kullback-Leibler divergence. Blue overlay represents standard deviation across disorder realizations.}\label{fig1}
\end{figure*}

Recently, it has been shown that deep learning algorithms can, in some cases, efficiently represent stationary quantum many-body states  \cite{levine2019quantum,gao2017efficient,carleo2017solving, noe2019boltzmann} or learn hard quantum distributions \cite{rocchetto2018learning}. Here we apply deep learning to analyze and simulate correlated quantum dynamics, focusing on correlated few-body quantum walks. To this end we employ a breakthrough in unsupervised machine learning: generative adversarial networks (GANs) \cite{goodfellow2014generative}. This deep learning method processes a dataset of examples (usually complex images) in an unsupervised manner using two neural networks that compete against each other to generate new, synthetic samples that are hard to distinguish from real ones \cite{TPDNE}. The Style Generative Adversarial Network (StyleGAN) extends the GAN architecture \cite{karras2019style,karras2020analyzing}, suggesting a model to disentangle and control different high-level properties in the generated images. Such models can generate impressively realistic images of (for example) faces, while offering computation-free control over high-level semantic features in the generated images, such as gender, age, facial expression, etc.

In this Letter, we show that when applied to physical problems (specifically, correlated quantum walks on disordered lattices), GANs can capture physical rules from examples only, without supervision. After training on a set of examples, they can generate without further calculations a much larger number of synthetic, unseen yet physically correct samples. Most interestingly, they automatically identify, and allow tuning of, physically meaningful variables (disorder level in our case) in the generated samples . 

\emph{Training the GAN algorithm.--} 
We consider two indistinguishable boson particles, propagating on one-dimensional lattices with ten sites, disordered on-site energies and uniform nearest-neighbor hopping. The two-body Bose-Hubbard Hamiltonian can be constructed from the lattice parameters: 
\begin{equation*}
H=\sum_{m}E_{m}a_{m}^{\dagger}a_{m}+\sum_{\langle n,m\rangle}J_{n,m}a_{n}^{\dagger}a_{m}+\frac{\Gamma}{2}\sum_{m}\hat{n}_{m}(\hat{n}_{m}-1)\label{eq:QH-1}
\end{equation*}
where $E_{m}$ is the on-site energy of site $m$, $a_{m}^{\dagger}\backslash a_{m}$ are the creation\textbackslash{}annihilation operator for a particle at site $m$, and $J_{n,m}\leq0$ is the hoping rate between neighboring lattice sites. $\hat{n}_{m}=a_{m}^{\dagger}a_{m}$ is the corresponding number operator, and $\Gamma$ is the on-site interaction energy between particles. In all the following cases, the parameters  $J_{n,m}=-1$ and $E_m$ are randomly chosen from a uniform distribution $0<E_m<3$. The disorder level is chosen such that the localization length is of the order of the lattice size, and therefor the propagating particles do not remain tightly localized at their starting position  \cite{lahini2008anderson}. 

To generate the training data set, we first consider a specific initial condition in which the two particles are placed at two adjacent sites at the center of the lattice 
$\left|\psi_{initial}\right\rangle=a_{5}^{\dagger}a_{6}^{\dagger}\left|0\right\rangle $. To generate the training samples, we use exact diagonalization to numerically propagate the initial state according to the Hamiltonian for a time $t=2$, so that the propagating particles do not reach and reflect from the edges of the lattice. Next, we calculate the two-particle correlation $\gamma_{q,r}=\left\langle a_{q}^{\dagger}a_{r}^{\dagger}a_{q}a_{r}\right\rangle $ which represents the probability of finding exactly one particle at site $q$ and one particle at site $r$ at the end of the evolution, taking into account the interference of all possible two particle processes, the quantum statistics, the interaction between the particles and the effect of the disordered lattice potential (see Fig 1b). Similar calculations were used previously by some of us to study correlated two-particle quantum walks on periodic lattices, in both simulations and experiments using photons \cite{bromberg2009quantum, peruzzo2010quantum} and ultra-cold atoms \cite{lahini_2particles_PhysRevA.86.011603,preiss2015strongly}. It was found that the choice of initial state and the level of interactions change the correlation map significantly. Here we study two-body and three-body quantum walk on disordered lattices.  

The GAN algorithm processes images, which here are prepared in the following way: for each random realization of the lattice parameters, we generate a single training image, an example of which is shown in the Supplementary Material \cite{Supplement}. Each image has two parts. The left panel of the image contains a column of ten pixels, representing the specific random realization of the on-site energies $E_{m}$, where the color code represents the value of each entry in 8-bit RGB. The right side of the training image represents the calculated two-particle correlation at the end of the propagation, as explained above.  We generate three separate data sets for three levels of interactions: $\Gamma=0$ for the interaction free case, $\Gamma=3$ for intermediate interactions and $\Gamma=1000$ for strong interactions – corresponding to hard-core bosons \cite{lahini_2particles_PhysRevA.86.011603,preiss2015strongly}. Each training set contains 70,000 training images, each with a different realization of disorder. The performance during learning is evaluated using the Frechet Inception Distance (FID) score  \cite{heusel2017gans}, which is often used to evaluate GANs images' quality. Additional information on the GAN architecture used here is detailed in the Supplementary Material \cite{Supplement}.

\begin{figure}
\includegraphics[width=0.49\textwidth]{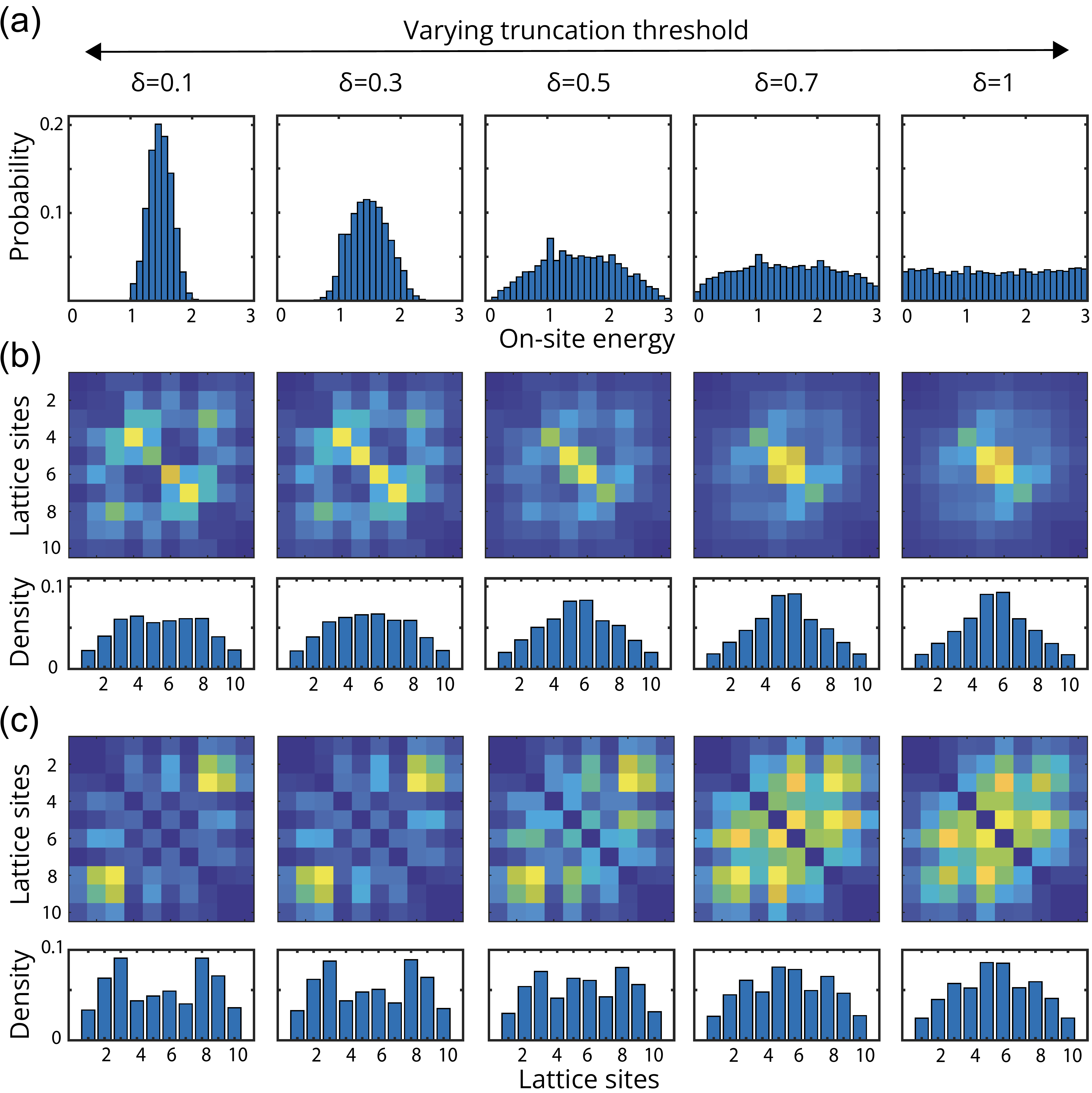}
\vspace*{-0.3cm}
\caption{\textbf{Latent-space truncation trick tunes the level of disorder in the generated samples. (a)} histograms of generated on-site energies from 1000 samples as a function of the truncation parameter $\delta$. While the original network was trained on disorder level corresponding to the rightmost histogram, truncating the latent space produces different levels of disorder. \textbf{(b)} Disorder-averaged correlation functions (top) and densities (bottom) for two particles initially on the same site and zero interaction. \textbf{(c)} Disorder-averaged correlation functions (top) and densities (bottom) for two particles initially on adjacent site an $\Gamma=3$. All results are physically valid with Kullback-Leibler divergence below 0.03.}\label{fig2}
\end{figure}

\emph{Generating new, unseen correlated quantum walks.--} 
After training is complete, the trained networks are used to generate new, synthetic samples, which look similar to the training images. To test if the synthetic images represent new and valid physical results, we extract from each generated image the values of the on-site energies $E_{m}$, calculate the expected two-body correlation after propagation using exact diagonalization, and compare the resulting correlation matrix to the generated one. The similarity between the generated and the exact correlations is evaluated using the Kullback–Leibler divergence $ KL=\sum P\left( x_1,x_2\right) \log \left( \frac{P\left( x_1,x_2\right) }{Q\left( x_1,x_2\right) }\right) $ were the reference distribution $P\left( x_1,x_2\right)$ is the exact normalized correlation function and the tested distribution $Q\left( x_1,x_2\right)$ is the normalized generated distribution. Figure 1 (c)-(d) show the quality of generated samples as the learning progresses. Initially the GAN generates blurred images with corresponding high $KL$ score, yet after learning the trained network generates synthetic data of random on-site energies and two-particle correlation with $KL<0.025 \pm 0.015$. In comparison, random normalized test distributions yield $KL>0.5 \pm 0.09$. This result means that after training, the GAN network can produce, without further calculations, new pairs of random lattice parameters and resulting two-particle correlations with the correct physics. The number of possible generated realizations is only limited by the random seed size, $2^{32}$ in our case. This indicates that the algorithm can correctly interpolate a cloud of examples to create a denser data-set, without explicitly calculating it. 

\emph{Latent space truncation tunes disorder level.--} 
Next, we show that although unsupervised, the algorithm identifies disorder as the relevant control parameter. As a result, the algorithm can be tuned, post-training, to generate on demand synthetic samples with different levels of disorder. To this end we use the "truncation trick" \cite{brock2018large} used in the StyleGAN algorithm \cite{karras2019style,karras2019analyzing}. In this procedure the intermediate latent space of the generator is truncated, usually to deal with areas that are poorly represented in the training data.

The training set used here has a fixed level of disorder, i.e the values of the lattice on-site energies $E_{m}$ are randomly chosen from the interval $0<E_{m}<\eta$ with $\eta=3$. We find that by controlling the truncation threshold $\delta$, the generator can be made to produce synthetic samples with a controlled level of disorder, i.e. with $\eta$ between 0 and 3. This is demonstrated in Figure 2 (a), which displays the histogram of generated $E_{m}$  values over 1000 generated samples (total of 10,000 values), as a function of the truncation parameter $\delta$. As $\delta$ is varied between 0 and 1, the spread of the histogram grows, reaching full spread (similar to the training set) at $\delta=1$. Thus, as $\delta$ is increased the level of disorder in the generated samples increases as well. As a result, the correlation matrices and particle densities exhibit a crossover from free propagation to Anderson localization \cite{lahini2008anderson}. Notably, this procedure does not degrade the fidelity of the generated correlation matrices and density distribution, i.e. the results remain physically valid with Kullback–Leibler divergence below 0.03 in all cases.

\emph{Generating unitary transformations.--}
So far, we have restricted ourselves to unsupervised learning of two-particle correlation functions and to a particular initial state – the two particles are initially on adjacent sites at the center of the lattice. Taking this technique a step forward, we consider now the full two-particle problem, i.e. considering all possible initial states and calculating the resulting complex-valued two-particle wave function. To this end, we construct the two-body Hamiltonian in the occupation number basis. The size of this real and symmetric matrix, $55\times55$, reflects the size of the Hilbert space, or the number of ways in which two particles can be arranged on ten sites. The unitary propagator, $U=e^{-iHt}$ is a complex valued and symmetric matrix of similar size, whose columns represent the resulting two-particle wavefunction, in the number basis, for each initial arrangement. Here, the training images contain three parts as shown in the Supplementary Information \cite{Supplement}. Here, the left panel of the image represents the ten value of the random on-site energies $E_{m}$ as before. The right panel of the training image is a matrix $55\times55$ which is divided in two. The bottom left half encodes the magnitude of the resulting unitary $U$, while the upper right half encodes its phase. 

As before, we generate three separate training sets for three levels of interactions  $\Gamma=0, 3, 1000$, each containing 70,000 training images with different disorder realizations. The GAN network is then trained on each dataset, while the performance during learning is evaluated using the Frechet Inception Distance score.  The training here takes longer, as expected, but by the end the trained network can generate synthetic novel samples. The physical validity of the synthetic samples was tested by extracting the generated on-site energies $E_{m}$, calculating the two-body Hamiltonian, exponentiating to obtain the expected unitary transform $U_{e}$, and comparing it to the generated unitary $U_g$. The fidelity was calculated using the standard form,  $Fidelity=\left|\frac{\textit{trace}(U_{e}^\dagger U_g)}{N}\right|$. Fig 3a shows the quality of generated samples as the learning progresses. At the end of training the network generates synthetic data of random on-site energies and unitary evolution with fidelity surpassing 92\%. Further details, as well as examples for the three-particle problem are presented in the Supplementary information \cite{Supplement}. 

Next, we show that here as well, latent space truncation results in post-training tuning of the level of disorder. Results are presented in Figure 3 (b), where in all cases the fidelity remains similar to the one obtained for the trained level of disorder, above 85\%.

\begin{figure}
\includegraphics[width=0.45\textwidth]{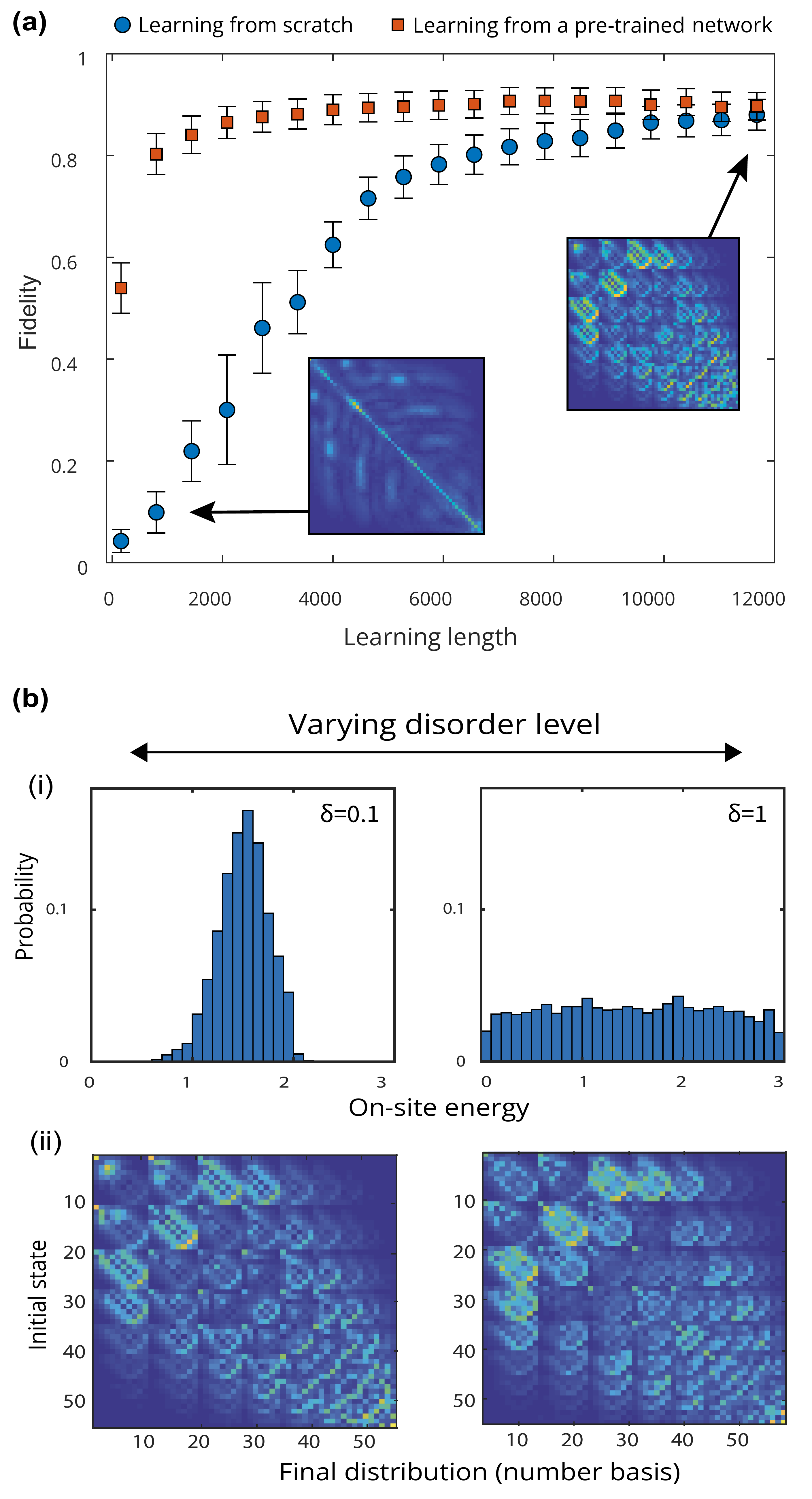}
\caption{Learning and tuning full unitary transformations of correlated quantum walks on disordered lattices. (a) Fidelity of the generated samples during learning, for a GAN network starting from scratch (circles) and for a pre-trained network (squares). Insets show examples of disorder-averaged unitaries at the beginning and the end of training. (b) Latent-space truncation used to tune disorder level in the generated two-particle unitaries.(i) histograms of generated on-site energies from 1000 samples as a function of the truncation parameter $\delta$. (ii) Disorder-averaged two particles unitaries (magnitude).}\label{fig5}
\end{figure}

\emph{Training using a pre-trained network.--} 
Finally, we show that the training time can be significantly shortened by taking as the starting point a network that completed training on different regime of parameters. Figure 3 (a) shows the training dynamics of two networks, both learning a system of two particles on ten sites and interactions level $\Gamma=1000$. The starting point for training in the two cases is different: while one network started from scratch (blue circles), the starting point for the second network is a network that completed training on the data set for $\Gamma=3$ (red squares). Notably, although the interaction levels are different, the second network has been able to optimize learning and use the knowledge accumulated in learning a different (yet related) problem. Indeed, the training of the second network converge much faster.

\emph{Discussion.--}
In this work we have shown that a machine learning algorithm based on GAN can learn the physical rules of correlated few-body quantum dynamics on disordered lattices in a completely unsupervised manner. The trained network can produce, without further calculation, new, unseen, and physically valid samples containing both lattice parameters and the resulting correlation functions or full unitary evolution. 

Remarkably, the GAN algorithm automatically identifies the relevant physical control parameter - disorder level in our case. This information is encapsulated in the algorithm's latent space; truncating it results in computation-free tuning of the disorder level in the generated correlated quantum walks, even though the network was not trained on these ranges explicitly. 

Recent advances in the applications StyleGAN algorithms to real-life images, (e.g. faces), demonstrated the ability to disentangle and separately tune multiple high-level semantic features in the generated images - for example age, gender, mood, hair length etc \cite{wu2021stylespace, pmlr-v119-voynov20a}. Our results suggest that when applied to physical problems, similar latent space operations may allow automated discovery and separate tuning of multiple real physical parameters, enabling computation-free exploration of complex physical problems. 

While here we focused on quantum dynamics problems and simulated training datasets, our algorithm could be used on a range of other physical problems and experimentally obtained data. This approach could become a tool for identifying relevant control parameters in complex data sets. It could also be used to interpolate sparse data sets correctly and quickly with generated data.

\emph{Acknowledgments.--} This research was supported by the Israel Science Foundation (ISF) and the Directorate for Defense Research and Development (DDR\&D) grant No. 3427/21 and by the Israeli Ministry of Science and Technology Grant No. 3-15671.

\bibliography{QW_bib}
  
\end{document}